\documentclass[aps,prl,twocolumn,superscriptaddress,showpacs,10pt]{revtex4-1}
\usepackage[utf8]{inputenc}
\usepackage{amssymb}
\usepackage{amsmath}
\usepackage{color}
\usepackage{dsfont}
\usepackage{graphicx}
\usepackage{braket}
\usepackage{hyperref}
\newcommand{\JH}{\ensuremath{J_\mathrm{H}}}

\newcommand{\ttg}{\ensuremath{{t_\mathrm{2g}}}}
\newcommand{\awg}{\ensuremath{{a_\mathrm{1g}}}}
\newcommand{\egp}{{\ensuremath{e_\mathrm{g}^\prime}}}

\newcommand{\LVO}{LiV$_2$O$_4$}

\begin{document}
\title{
\textit{Ab initio} study on heavy-fermion behavior in {\LVO}:\\
Role of Hund's coupling and stability
}
\author{Steffen Backes}
\email{steffen.backes@riken.jp}
\affiliation{RIKEN iTHEMS,  Wako, Saitama 351-0198, Japan}
\affiliation{Center for Emergent Matter Science, RIKEN, Wako, Saitama 351-0198, Japan}

\author{Yusuke Nomura}
\affiliation{Institute for Materials Research, Tohoku University, Sendai 980-8577, Japan}

\author{Ryotaro Arita}
\affiliation{Department of Physics, University of Tokyo, Hongo Bunkyo-ku Tokyo 113-0033 Japan}
\affiliation{Center for Emergent Matter Science, RIKEN, Wako, Saitama 351-0198, Japan}

\author{Hiroshi Shinaoka} 
\affiliation{Department of Physics, Saitama University, Saitama 338-8570, Japan}

\date{\today}

\begin{abstract}
{\LVO} is a member of the so-called $3d$ heavy fermion compounds, with effective electron mass exceeding 60 times the free electron mass, comparable to $4f$ heavy fermion compounds. 
The origin of the strong electron correlation in combination with its metallic character have been a subject of intense theoretical and experimental discussion, with Kondo-like physics and Mott-physics being suggested as its physical origin.
Here we report state-of-the art \textit{ab initio} Density Functional Theory + dynamical mean-field theory calculations for {\LVO} for the full three orbital V {\ttg} manifold, and present temperature-dependent spectral properties. We map out the phase diagram for a representative 3-orbital model system as a function of doping and interaction strength, which indicates that {\LVO} is located between two orbital-selective Mott phases, giving rise to a robust strongly correlated Hund's metal behavior. At low temperature we find the emergence of a strongly renormalized sharp quasi-particle peak a few meV above the Fermi level of V {\awg} character, in agreement with experimental reports.
\end{abstract}


\maketitle

\paragraph{Introduction.}
The transition metal oxide {\LVO} is the first heavy fermion (HF) compound discovered which valence electron states are not composed of $f$-electrons but $3d$ states~\cite{Kondo1999,Anisimov:1999jz,Johnston1999,TAKAGI1999147,2012NatCo...3E.981S,JOHNSTON200021}. The Sommerfeld coffecient at low temperature reaches $\gamma(1K)\approx 0.42$~J/mol K$^2$~\cite{MATSUNO:1999th,Johnston1999}, which is an exceptionally large value for a metallic $d$-electron compound. Below a characteristic temperature of $T\approx 20-30$~K~\cite{Kondo:1997kp,Johnston1999,JOHNSTON200021}, {\LVO} displays strongly enhanced electronic specific heat $\gamma$ and a $T^2$ dependence of the electrical resistivity, characteristic of HF behavior with the emergence of Fermi-liquid behavior at low temperatures~\cite{Kondo:1997kp,TAKAGI1999147,Urano2000,Lee2001}
photoemission spectroscopy experiments (PES) have found the emergence of a sharp peak structure in the spectral function $\sim 4$~meV above the Fermi level~\cite{Shimoyamada2006}, which gradually vanished as temperature is increased, similar to a Kondo resonance peak found in other heavy fermion compounds~\cite{Lee1986,Allen2005,Strydom2023}.

Several theoretical explanations have been proposed to explain this unique behavior. Band structure calculations show that three bands derived from V {\ttg} orbitals cross the Fermi level: a doubly degenerate {\egp} band with bandwidth $\sim 2$~eV, and a nondegenerate {\awg} band with bandwidth $\sim 0.2$~eV~\cite{Eyert1999,Singh1999}. Closeness to a Mott insulating phase has been suggested~\cite{TAKAGI1999147}, as well as Kondo-lattice like behavior from itinerant {\egp} electrons hybridizing with the localized {\awg}, thus identifying the spectral feature at $\approx 4$~meV above the Fermi level as a Kondo resonance~\cite{Anisimov:1999jz}.
It was further suggested that the geometrical frustration of the V pyrochlore sublattice could lead to strong spin and orbital fluctuations~\cite{Eyert1999,Urano2000,Fulde2001,Fujimoto2002,Hopkinson2002}.
Density Functional Theory + dynamical mean-field theory (DFT+DMFT) calculations have indicated that {\LVO} can be considered as a slightly doped Mott-insulator~\cite{Nekrasov:2003kj,Arita:2007bo}, ruling out Kondo physics from hybridization of the {\awg} with the dispersive {\egp} orbitals as the origin of the strong correlations~\cite{Arita:2007bo}.
While these works studied the effect of strong Coulomb interactions in a simplified two-orbital model~\cite{Arita:2007bo} or
with neglected {\awg}-{\egp} hybridization above the coherence temperature~\cite{Nekrasov:2003kj},
possible Hund's physics~\cite{Yin2011,Medici2011,Yin2011_2,Villar2021} induced by {\JH} in the full V {\ttg} manifold of {\LVO} and robustness with respect to interaction strength and doping have not been investigated.

Here we report \textit{ab initio} DFT+DMFT calculations for the low-$T$ heavy-fermion state with the full 3 orbital {\ttg} manifold, and \textit{ab initio} derived Coulomb interaction parameters. We perform a systematic study of the electronic correlation strength, spectral function, and the dependence on interaction parameters, electronic filling and temperature, in order to map out the phase diagram of {\LVO} as a function of interaction strength and doping.
We find signatures of a strongly correlated Hund's metal, with the role of Hund's coupling {\JH} being two fold:
First, {\JH} favors a localized moment in the narrow  V {\awg} orbital and, second, it induces a robust heavy fermion metal behavior even in the "substantially" doped case. The heavy fermion phase persists in a wide range of interaction strength $U$ and doping,
stabilized by the multi-orbital character and band-width differentiation of the V {\ttg} manifold.
We observe the emergence of a sharp peak in the spectral function above the Fermi level only at low temperatures, in agreement with experimental observations.


\begin{figure}[t]
	\centering
	\includegraphics[width=.99\linewidth,clip]{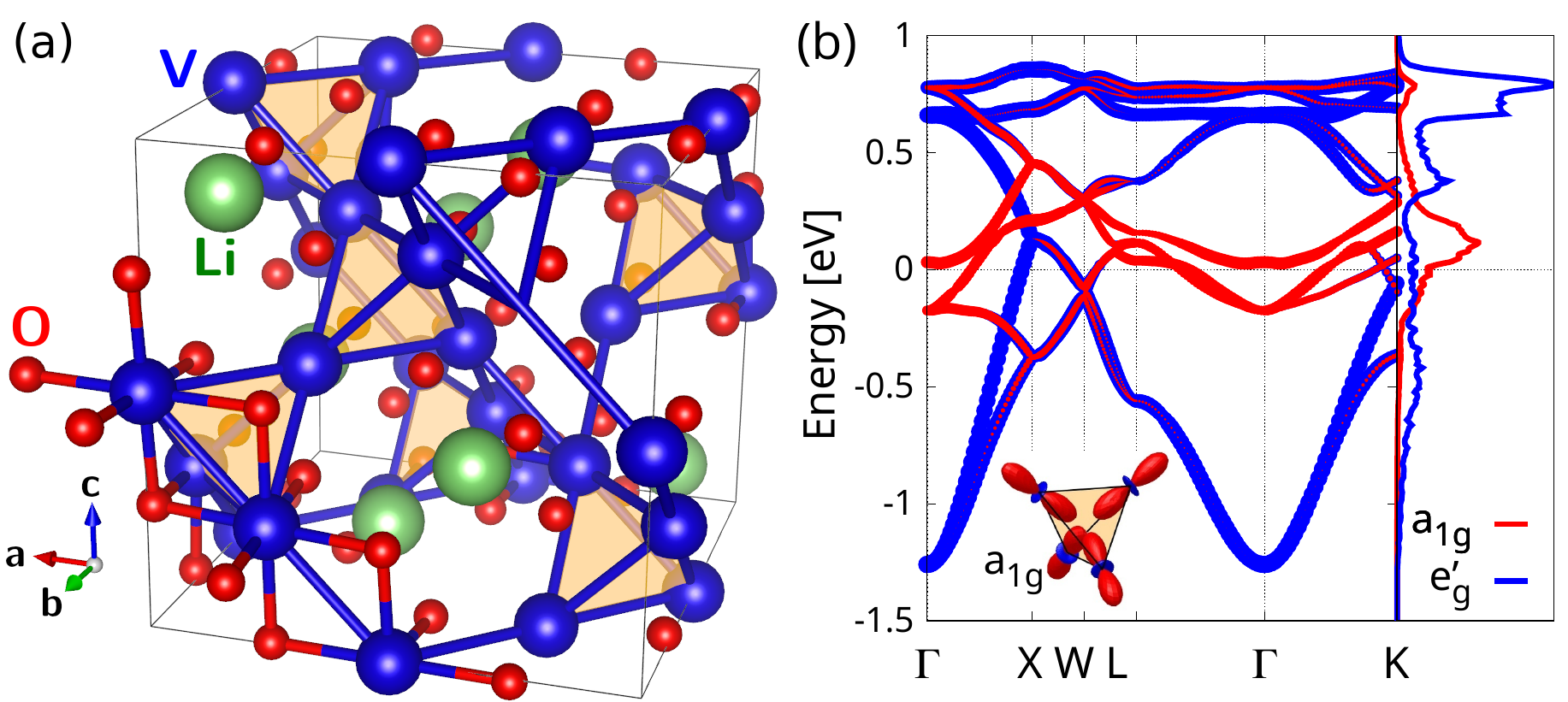}
	\caption{
	(a) Crystal structure {\LVO}. The vanadium atoms (blue) form a frustrated pyrochlore lattice, and each V atom is surrounded by a VO$_6$ tetrahedron with slight trigonal distortion. For better visibility only a few V-O bonds are indicated.
	(b) Density-functional theory-derived band structure and density of states close to the Fermi level, with the colors indicating orbital character. The V {\ttg} manifold consists of two degenerate {\egp} and one {\awg} orbital (see inset), well separated from other states at higher energies.
	}
	\label{fig:crystal-structure}
\end{figure}

\paragraph{Methods.}
We construct an \textit{ab initio} model for {\LVO} by obtaining maximally localized Wannier functions for the V $t_\mathrm{2g}$ orbitals, based on Density Functional Theory (DFT) as implemented in \texttt{Quantum ESPRESSO}~\cite{Giannozzi_2009,Giannozzi_2017} and \texttt{wannier90}~\cite{Wannier90}, using the LDA exchange-correlation functional~\cite{Ceperley:1980zz,Perdew:1981dv}.
The experimental crystal structure at 4 K~\cite{Chmaissem:1997df} was used (see Fig.~\ref{fig:crystal-structure} (a)),
where each V atom is located at the center of an O octahedron, slightly distorted along its local [111] axis (trigonal distortion).
This gives rise to a degeneracy lifting of the
$d_{xy}$--, $d_{yz}$--, $d_{zx}$-- orbitals, resulting in two-fold degenerate {\egp} orbitals and one {\awg} orbital.
Based on the DFT calculation, we use the constrained random-phase approximation (cRPA)\cite{respack} to calculate the effective Slater-Kanamori interactions
\begin{align}
	& H_\text{SK}(U, U^\prime, \JH) \nonumber \\
	&=
	\sum_\alpha U n_{\alpha\uparrow} n_{\alpha\downarrow}  + \sum_{\alpha<\beta,\sigma}
	\Big[
	U^\prime n_{\alpha\sigma} n_{\beta\bar{\sigma}}
	+(U^\prime - \JH) n_{\alpha\sigma}n_{\beta\sigma}
	\Big]\nonumber\\
	&+\JH\sum_{\alpha\neq\beta}\Big[
	c^\dagger_{\alpha\uparrow}
	c^\dagger_{\beta\downarrow}
	c_{\alpha\downarrow}
	c_{\beta\uparrow}
	+
	c^\dagger_{\alpha\uparrow}
	c^\dagger_{\alpha\downarrow}
	c_{\beta\downarrow}
	c_{\beta\uparrow}
	\Big]
\label{eq:Kanamori_interaction}
\end{align}
with $U$=3.91 eV, $U^\prime$ =2.8 eV, $\JH$ = 0.54 eV.
The Coulomb interaction is nearly isotropic with $U - 2\JH \simeq U^\prime$. Our values agree well with another recent study~\cite{grundner2024}.
Supplementing the non-interacting Wannier model with this interaction term, we perform self-consistent
DMFT calculations, as implemented in \texttt{DCore}~\cite{Shinaoka2021-iv}, using a continuous-time Quantum Monte-Carlo impurity solver in the hybridization expansion based on \texttt{ALPS/CT-HYB}~\cite{Gaenko:2017ic,Bauer:2011tz,SHINAOKA2017128}.
We neglect spin-orbit coupling, and solve the impurity problem in the eigenbasis of the crystal field splitting, which diagonalizes the local orbital levels and hybridization function, and does  not give rise to a sign problem. Spectral properties were obtained by performing analytic continuation on the imaginary frequency data using the Maximum Entropy method~\cite{Silver1990,Jarrell1996} implemented in \texttt{ALPS/CQMP}~\cite{Levy:2017ek,Gaenko:2017ic,Bauer:2011tz}, and Pad\'e approximation~\cite{Vidberg1977,Gunnarsson2010} close to the Fermi level.


\paragraph{Results - Local spectral function.}
Figure~\ref{fig:crystal-structure}(b) shows the computed LDA band structure, with the different colors indicating the weight of the {\awg} and {\egp} orbital character.
The $t_\mathrm{2g}$ manifold is separated from the rest of the bands, and we see that the {\awg} derived bands have a very small effective bandwidth of about $0.1-0.2$~eV, much smaller than the {\egp} derived bands ($\sim 2$~eV), as already pointed out in a previous study~\cite{Anisimov:1999jz}.

\begin{figure}[t]
 	\includegraphics[width=.99\linewidth,clip]{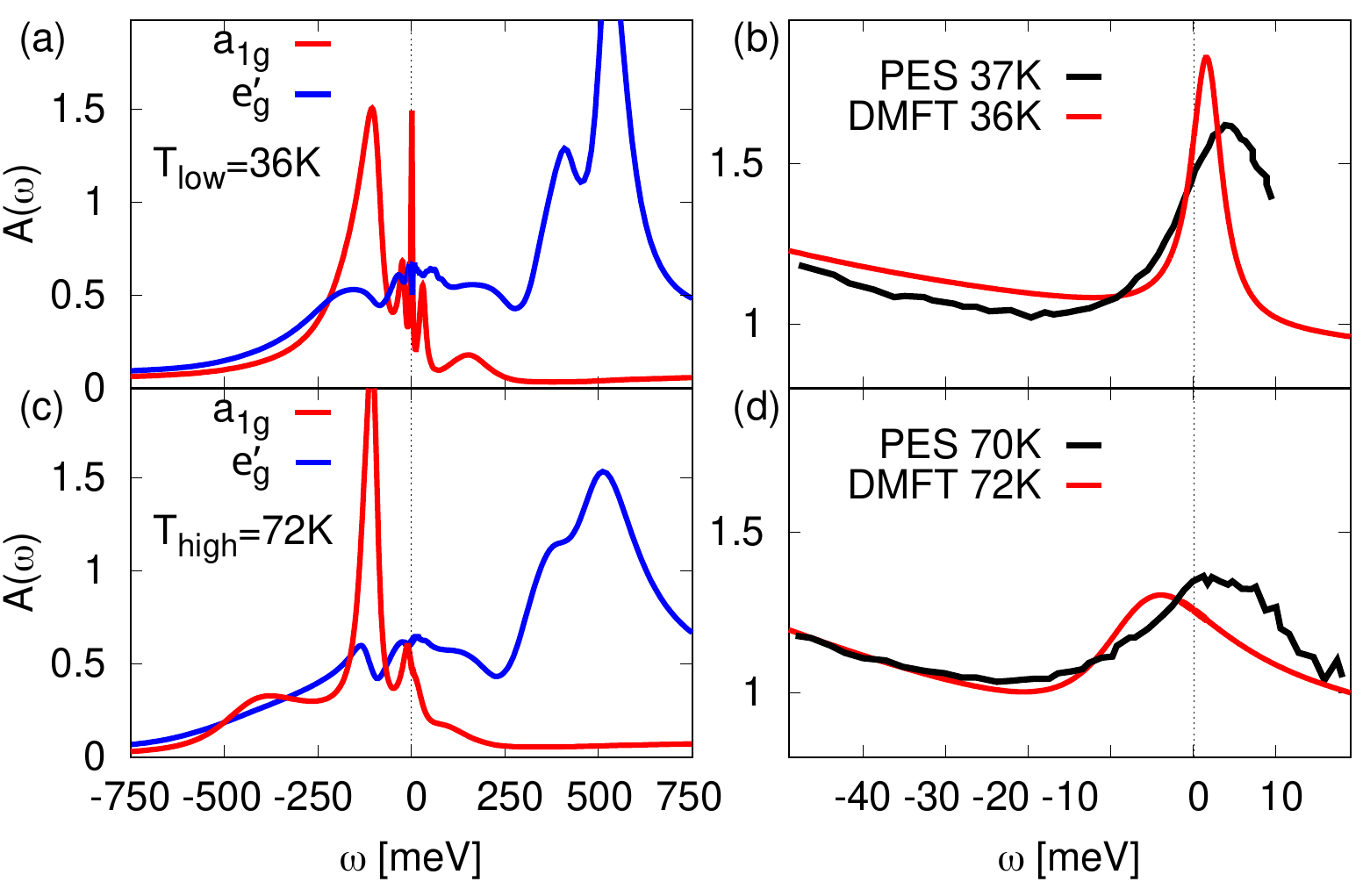}
	\caption{
	The DFT+DMFT orbital-resolved local spectral function $A(\omega)$ (a),(c) of {\LVO} for the two temperatures $T=36$~K and $T=72$~K.
	(b),(d) shows a close-up of the total spectral function around $E_F$, compared to PES data from~\cite{Shimoyamada2006}.
	At low temperature a sharp quasi-particle peak of V {\awg} character emerges just above the Fermi level.
	}
	\label{fig:locspec}
\end{figure}

We now study the effects of the local Coulomb interaction at low $T$ by means of DMFT calculations.
Figure~\ref{fig:locspec} shows the orbital-resolved local spectral function from DFT+DMFT.
We observe a strongly renormalized {\awg} orbital, with weakly correlated {\egp} orbitals.
Accordingly, we find the mass enhancement factor of the {\awg} orbital ($m^*/m\approx 58$) to be one order of magnitude larger than for the {\egp} orbital ($m^*/m\approx 5$ ).
Around $150$~meV binding energy the {\awg} orbital forms a Hubbard-band like peak in the spectral function, with no corresponding similar peak mirrored at positive energies, reflecting the absence of particle-hole symmetry.
At low temperature, a sharp quasi-particle peak of V {\awg} character emerges just above the Fermi level at around $\approx 2$~meV, which agrees well with experimental PES data (see Fig.~\ref{fig:locspec} (b)). At higher temperature, this peak is quickly suppressed and almost vanishes in our calculation at $72$~K, with a small hump of spectral weight remaining, which also appears to qualitatively agree with experimental observations. The {\egp} spectral function is weakly renormalized with considerable weight at the Fermi level, and stays essentially temperature independent.
Thus, our result for the full V {\ttg} manifold spectral function confirms the previous DMFT results from a simplified 2 orbital model~\cite{Arita:2007bo}.

Compared to the DFT density of states, we observe a significant shift of spectral weight in the {\awg} orbital to energies below the Fermi level. A charge of about $0.5$e is transferred from the {\egp} to the {\awg} orbital, as shown in the orbital occupations in Table~\ref{tab:orbital_fillings}. This leads to an {\awg} electron filling which is $10$\% hole-doped compared to half-filling ($0.45$e per spin orbital), while the {\egp} orbitals remain at a filling slightly larger than quarter filling ($0.3$e per spin orbital). This allows the system to lower its energy by partially localizing the electrons in the {\awg} orbital with the smaller bandwidth to avoid the Coulomb interaction energy penalty. The remaining itinerant electrons in the {\egp} orbitals lower the energy via the kinetic energy contribution from its large bandwidth. This picture is confirmed in the calculated spin-spin correlation function $\chi_{\rm loc}$ (see supplementary material), which shows the emergence of fluctuating local moments in the {\awg} orbital at low temperature.
We do not observe tendencies to form an orbital-selective Mott (OSM) phase with fully localized {\awg} electrons, which would technically be allowed in a $1+2$ degenerate orbital system away from integer electron filling. Though, we find the system to be close to an OSM transition when doping additional charge carriers, resulting in the observed strong correlation effects, as discussed further down below.

\begin{table}[t]
    \centering
    \begin{tabular}{|c|c|c|}
    \hline  orbital & DFT & DFT+DMFT \\
    \hline  {\awg}  & $0.41$e & $0.90$e ($+0.49$e)  \\
       {\egp}  & $1.09$e & $0.60$e ($-0.49$e) \\
	\hline Total      & $1.5$e & $1.5$e  \\      \hline
    \end{tabular}
    \caption{V {\ttg} orbital occupations (summed over spin degrees of freedom) as obtained from DFT and DFT+DMFT at $T=36$~K.
    Local Coulomb correlations lead to significant charge transfer ($\sim 0.5$e) from the
    two {\egp} orbitals into the {\awg} orbital. Occupations at $T=72$~K identical within $1\%$.
    }
    \label{tab:orbital_fillings}
\end{table}

\paragraph{Momentum resolved spectral function and Fermi surface.}
\begin{figure}[t]
	\includegraphics[width=.99\linewidth,clip]{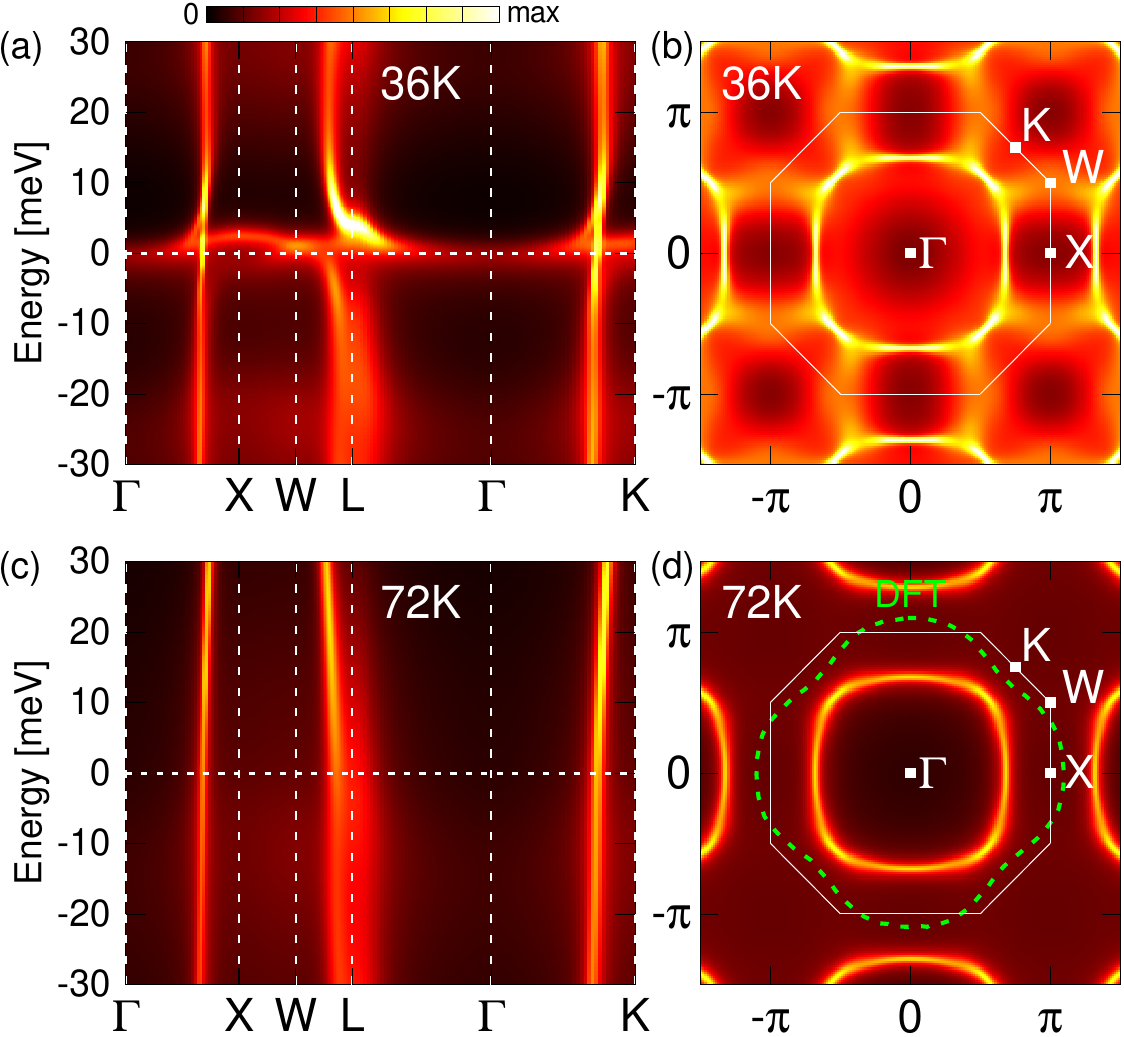}
	\caption{
	(a),(c): The DFT+DMFT momentum-resolved spectral function $A({\mathbf k }, \omega)$  of {\LVO} for the two temperatures $T=36$~K and $T=72$~K. The highly-renormalized {\awg} states appear at low temperature as an extremely flat but slightly dispersive excitation around the Fermi level. The {\egp} states form weakly correlated highly dispersive bands that show no significant temperature dependence. \\
	(b),(d): The Fermi surface in the $k_x,k_y$-plane.
	The {\awg} states appear at low temperatur as incoherent spectral weight around $K$, which vanishes at higher temperature.
	The {\egp} states create a coherent electron pocket around $\Gamma$, which is significantly smaller compared to the DFT pocket (green line, only first Brillouin zone shown) due to charger transfer into the {\awg} orbital.
	}
	\label{fig:FS_BS}
\end{figure}
In Fig.~\ref{fig:FS_BS} we show the obtained momentum-resolved spectral function $A({\mathbf k }, \omega)$  of {\LVO} for the two temperatures $T=36$~K and $T=72$~K, as well as the Fermi surfaces in the $k_x,k_y$-plane.
The sharp peak of {\awg} character that appears in the local spectral function at low temperature manifests itself as a flat, almost dispersionless state close to $E_F$, with spectral weight mostly located around the X,W, and K high-symmetry points. This state corresponds to the highly renormalized {\awg} orbital, which experiences a bandwidth reduction of about a factor of $50$, in accordance with its effective mass enhancement $m^*/m\approx 58$. The {\awg} spectral weight quickly becomes incoherent and suppressed when increasing the temperature, with no noticable weight remaining at $T=72$~K.
The {\egp} states form highly dispersive excitations with well-defined coherent electron pockets around the $\Gamma$ point and no significant temperature dependence is observed, apart from the vanishing hybridization with the temperature-dependent {\awg} states. Due to the charge transfer from the {\egp} into the {\awg} orbital the electron pocket size is found to shrink significantly compared to the DFT derived electron pocket.
These observations could in fact be indicative for a Kondo-type scenario, where the {\awg} local moments get screened by hybridizing with the dispersive {\egp} states at low temperature. Though, we will now show that {\LVO} can be more naturally understood as a multi-orbital Hund's metal, where the Hund's coupling {\JH} is the defining quantity that gives rise to the strongly renormalized {\awg} state.

\begin{figure}[t]
	\includegraphics[width=.99\linewidth,clip]{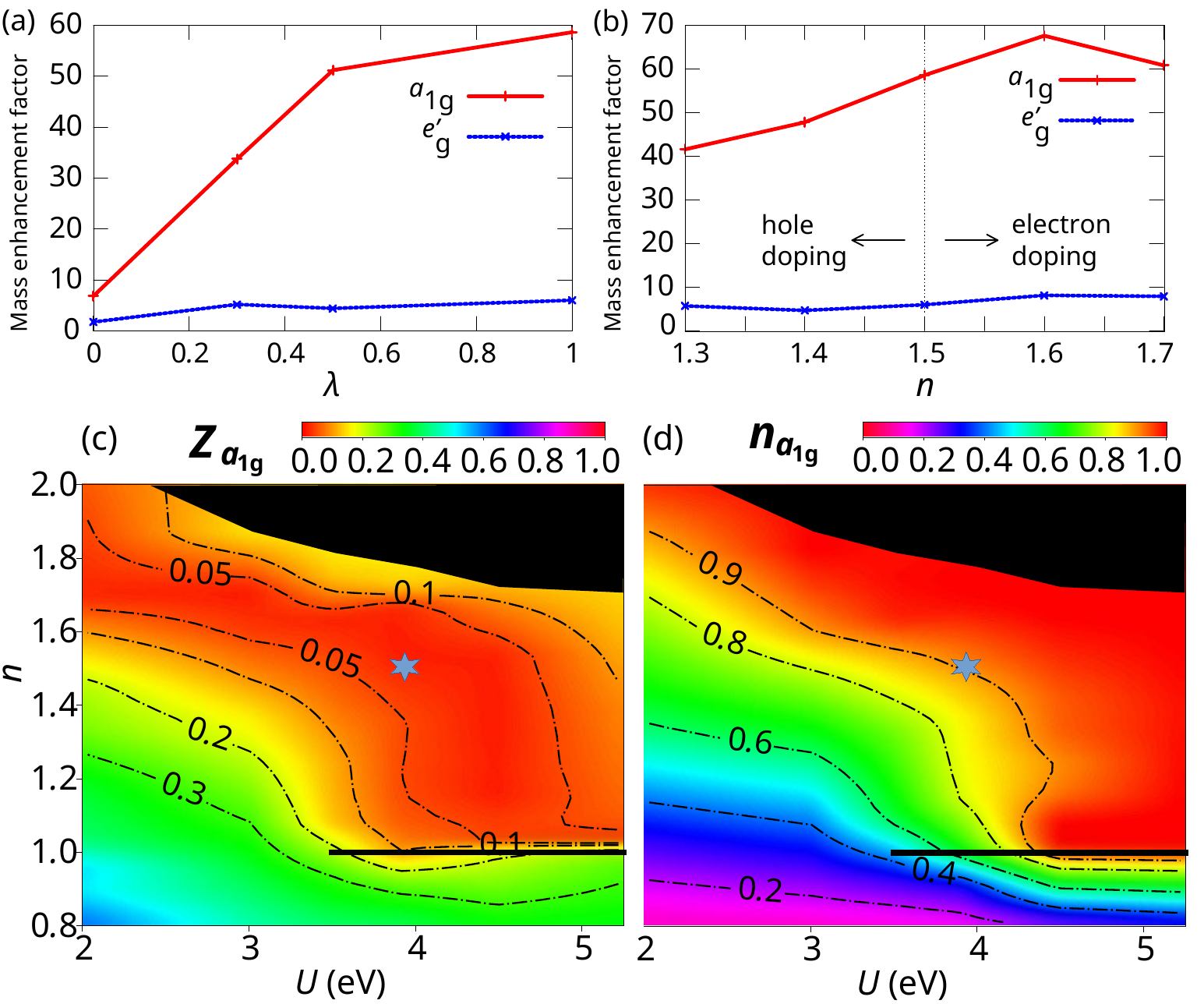}
		\caption{
		The electronic mass enhancement factor $m^*/m$ in {\LVO} calculated from DFT+DMFT at $T=24$ K as a function of
	(a) the Hund's coupling, scaled as $\lambda \JH$, $\lambda \in [0,1]$, 
	(b) the electron occupation.
	The mass enhancement is robust against charge or hole doping, and depends almost exclusively on the value of the Hund's coupling.
		(c)+(d) DMFT phase diagram of the 3-orbital model computed at $T=24$ K.
		(c) renormalization factor $Z$ for the $\awg$ orbital and (d) electronic occupation of $\awg$ orbital.
		The vertical axes denote the total electron number $n$. The Mott insulator transition at $n=1$ is of first order, and we show the result for a metallic starting point. The black area around $n=2$ indicates the half-filled {\awg} orbital-selective Mott phase.
		The parameters corresponding to {\LVO} are indicated by the blue star-shaped symbol (only approximate since the bandwidth are slightly different).
	}
	\label{fig:z-phasediag}
\end{figure}

\paragraph{Relevance of Hund's coupling {\JH}.}
To investigate the effects of $\JH$ on the electronic structure, 
we perform DMFT calculations while varying the strength of the Hund's coupling $\JH$ as $\lambda \JH$ with $\lambda \in [0,1]$.
Figure~\ref{fig:z-phasediag} (a) shows the mass enhancement factor $m^*/m$ as a function of $\lambda \JH$.
We observe a high sensitivity on the value of $\lambda \JH$, where the correlation strength is reduced by about an order of magnitude for $\lambda=0$ with $m^*/m<10$, but then $m^*/m$ for the {\awg} orbital rapidly increases to values over 50 already at $\lambda=0.5$.
Until the \textit{ab initio} value $\lambda=1$ it continues to rise only moderately, which indicates that {\LVO} is in fact deep in the Hund's metal regime. We would like to point out electronic correlation effects from \textit{Hubbard}/\textit{Mott} physics are still present at $\lambda \JH=0$, i.e. $U=U^\prime =3.91$~eV, where the system is only weakly correlated, despite the orbital-averaged interaction having increased. But only when $\JH$ is active, the system develops heavy fermion behavior, i.e. in fact \textit{Hund's} physics are the main driving force for the exceptionally strong correlations. $\JH$ promotes the formation of a high-spin configuration, which partially localizes the {\awg} electrons. This is also evident in the formation of a local moment observed in the spin-spin correlation function for increasing $\lambda$ (see supplementary material).
We find the large mass enhancement to persist for large range of interaction strength (see supplementary material for $U$ dependence, Fig.~\ref{fig:z-phasediag} (a) for $\JH$ dependence) and doping (see Fig.~\ref{fig:z-phasediag} (b)) as well, which shows that the HF state in {\LVO} should be very robust.

\paragraph{Simplified 3 orbital model and phase diagram.}
The large mass enhancement factor of $\ge 50$ in {\LVO} at a comparatively low filling of $n=1.5$ away from integer or half filling ($n=3$), combined with not being in immediate vicinity of a (orbital-selective) Mott transition is quite unusual. Mass enhancements of this scale in three orbital systems are usually found at comparatively large $U$ values or close to integer fillings\cite{Medici2011,Dang2015,Stadler2019,Deng2019,Stadler2021}.
To analyse the generality of the large mass enhancement, we construct a simplified three orbital model with the Hamiltonian
    $\mathcal{H} = \mathcal{H}_0 - \Delta~n_{\awg} + \mathcal{H}_\mathrm{SK}(U, U-2\JH, \JH),\label{eq:model-simplified}$
where the value of the crystal field $\Delta = 0.24$~eV is taken from the strength of the crystal field estimated by the Wannier functions for the compound. We fixed $\JH/U=0.14$ corresponding to {\LVO}.
We use a cubic lattice and set the full band widths to 3 and 1 eV for the $\awg$ and $\egp$ orbitals, respectively, and use the same form of the interaction as in Eq.~\eqref{eq:Kanamori_interaction}.
The resulting $U-n$ (interaction-filling) phase diagram obtained by DMFT at $T\simeq 24$ K is shown in Fig.~\ref{fig:z-phasediag} (c)-(d).
The {\awg} inverse mass enhancement factor $Z$ (Fig.~\ref{fig:z-phasediag} (c)) reveals two Mott-insulator lines at $n=1$ and $n=2$, but the latter one is completely covered by an orbital selective Mott (OSM) phase.
In the OSM phase, the insulating $\awg$ orbital is half-filled (black area), while the $\egp$ orbitals remain partially filled.
The recovery of $Z$ to larger values around $n=2$ is related to the absence of particle-hole symmetry and the self-energy peak being not centered the Fermi level, but we confirmed the insulating nature via the spectral weight at the Fermi level.
The Mott insulator at $n=1$ (black line) is of single-orbital nature,
being characterized by a half-filled $\awg$ orbital and unoccupied $\egp$ orbitals.
As the main result we find that the mass enhancement factor is substantially enhanced between these distinct Mott phases, forming a \textit{heavy-fermion (HF) phase}. This HF phase turns out to be robust and extends over a large range of dopings and interaction strengths, where the HF behavior is insensitive to changes in these parameters. Though, in many real materials this phase might be inaccessible because a symmetry-broken phase is favored, in {\LVO} geometrical frustration from the pyrochlore lattice prevents symmetry breaking and can stabilize the HF phase.
This  phase coincides with the region defined by $n_\awg > 0.8$, as can be seen in Fig.~\ref{fig:z-phasediag} (d). 
This area is quite unusual, since $n_\awg \sim 1$ even though $n_{\rm tot}$ is increased from $1$ to $2$. For $U \gtrsim 4$~eV, all additional charge carriers dope into the dispersive {\egp} orbitals, while the {\awg} orbital stays essentially invariant $n_\awg \sim 1$, giving rise to a robust HF phase for a large range of parameters. This is very different from degenerate multi-orbital systems, where the orbital fillings are identical by symmetry, and such orbital-differentiated HF phase for a wide range of doping cannot emerge. Thus, the large bandwidth differentiation between the {\egp} and {\awg} in {\LVO} is essential for hosting the robust HF phase.
The doping and interaction parameters corresponding to {\LVO} locate the system right in this HF phase, which explains why the exceptionally large mass enhancement was found to be stable against variations of the \textit{ab initio} parameters. While this behavior is similar to other Hund's metals\cite{Yin2011,Medici2011,Yin2011_2,Villar2021}, in {\LVO} the small {\awg} bandwidth due to the trigonal crystal field, the multi-orbital nature that allows for charge redistribution with the Hund's coupling inducing a local moment combine to create an exceptionally strongly correlated Hund's metal, with effective masses similar to $f$-electron systems.

\paragraph{Conclusions.}
In summary, we have studied the $3d$ heavy fermion compound {\LVO} using \textit{ab initio} DFT+DMFT calculations treating the full V {\ttg} manifold, going beyond previous DMFT investigations of simplified models. We obtained the local and momentum resolved spectral function and Fermi surface, and have identified the emergence of a sharp spectral peak of V {\awg} just above the Fermi level at low temperatures. This peak corresponded to a quasi-particle like excitation of the highly renormalized {\awg} orbital with mass enhancement $m^*/m > 50$, and was found to be in good agreement with photoemission spectroscopy data.
We confirmed the robustness of the strong electronic correlations with respect to a variation of the system's parameters, and mapped out the interaction-doping phase diagram for a representative three orbital model. From these results we found that {\LVO} can be considered as a strongly correlated Hund's metal, that is located in a HF phase right between two Mott-insulating states at higher doping. 
The trigonal crystal field splitting resulting in a narrow {\awg} and broad {\egp} orbitals allows for the unusual case where the {\awg} stays close to half-filling even with significant doping, and thus can host a robust heavy fermion phase insensitive to moderate changes of the system parameters.
Thus, we conclude that two criteria are essential in {\LVO} to realize the large mass enhancement and spectral peak: (1) the splitting into narrow {\awg} and broad {\egp} orbitals, allowing (2) multi orbital Hund's physics (rather than Kondo physics) to create an orbital selective robust heavy fermion phase. 
Our results have general implications for the understanding of heavy fermion materials and suggest further investigations of the role of the Hund's coupling in these materials.

During the final preparation stages of this manuscript we became aware of another preprint that investigated {\LVO} with similar methods~\cite{grundner2024}. While our general results agree, our work in addition presents the Fermi surface, comparison with experiment, and investigates the generality of the heavy fermion phase via the phase diagram of a representative three orbital model.

\paragraph{Acknowledgments.}
We thank S. Biermann, Y. Motome, H.  Takagi, A. Uehara, and P. Werner for fruitful discussions,
   and K. Held for helpful comments.
	We gratefully acknowledge support by the wider ALPS community~\cite{Bauer:2011tz,Albuquerque:2007ja}.
	This work was supported by JSPS KAKENHI Grant Number 15H05885 (J-Physics), 16K17735, and
    by the RIKEN TRIP initiative (RIKEN Quantum, Advanced General Intelligence for Science Program, Many-body Electron Systems).
    Y.N. acknowledges support from MEXT as ``Program for Promoting Researches on the Supercomputer Fugaku'' (Grant Number JPMXP1020230411) and JSPS KAKENHI (Grant Numbers JP23H04869, JP23H04519, and JP23K03307).
	Part of the calculations were performed on the ISSP supercomputing system.
	We used \texttt{VESTA}~\cite{Momma:2011dd} for visualizing the crystal structure.


%

\end{document}